\documentclass[12pt,preprint]{aastex}
\usepackage{natbib}
\usepackage{url}
\usepackage{amsmath}
\usepackage{multirow}
\usepackage{color}
\usepackage{ulem}

\shorttitle{MAGNETIC HELICITY AND CMES}
\shortauthors{PARK et al.}

\begin{document}

\title{THE OCCURRENCE AND SPEED OF CMES RELATED TO TWO CHARACTERISTIC EVOLUTION PATTERNS OF HELICITY INJECTION IN THEIR SOLAR SOURCE REGIONS}

\author{SUNG-HONG PARK\altaffilmark{1}, KYUNG-SUK CHO\altaffilmark{2,3,1}, SU-CHAN BONG\altaffilmark{1}, PANKAJ KUMAR\altaffilmark{1}, JONGCHUL CHAE\altaffilmark{4}, RUI LIU\altaffilmark{5,6}, AND  HAIMIN WANG\altaffilmark{5}}

\altaffiltext{1}{Korea Astronomy and Space Science Institute, Daejeon, 305-348, Republic of Korea; freemler@kasi.re.kr}
\altaffiltext{2}{NASA Goddard Space Flight Center, Greenbelt, MD 20771, USA}
\altaffiltext{3}{Department of Physics, The Catholic University of America, Washington, DC 20064, USA}
\altaffiltext{4}{Astronomy Program, Department of Physics and Astronomy, Seoul National University, Seoul, 151-742, Republic of Korea}
\altaffiltext{5}{Space Weather Research Laboratory, New Jersey Institute of Technology, Newark, NJ 07102, USA}
\altaffiltext{6}{CAS Key Lab of Geospace Environment, Department of Geophysics \& Planetary Sciences, University of Science \& Technology of China, Hefei, Anhui 230026, China}

\begin{abstract}
Long-term (a few days) variation of magnetic helicity injection was calculated for 28 solar active regions which produced 47 CMEs to find its relationships with the CME occurrence and speed using $SOHO$/MDI line-of-sight magnetograms. As a result, we found that the 47 CMEs can be categorized into two different groups by two characteristic evolution patterns of helicity injection in their source active regions which appeared for $\sim$0.5--4.5 days before their occurrence: (1) a monotonically increasing pattern with one sign of helicity (Group A; 30 CMEs in 23 active regions) and (2) a pattern of significant helicity injection followed by its sign reversal (Group B; 17 CMEs in 5 active regions). We also found that CME speed has a correlation with average helicity injection rate with linear correlation coefficients of 0.85 and 0.63 for Group A and Group B, respectively. In addition, these two CME groups show different characteristics as follows: (1) the average CME speed of Group B (1330$\,$km$\,$s$^{-1}$) is much faster than that of Group A (870$\,$km$\,$s$^{-1}$), (2) the CMEs in Group A tend to be single events, whereas those in Group B mainly consist of successive events, and (3) flares related to the CMEs in Group B are relatively more energetic and impulsive than those in Group A. Our findings therefore suggest that the two CME groups have different pre-CME conditions in their source active regions and different CME characteristics.
\end{abstract}


\section{Introduction}
\label{sec1}
Coronal Mass Ejections (CMEs) are transient ejections into interplanetary space of as much as about 10$^{13}$--10$^{17}\,$g of plasma and embedded magnetic fields from the solar corona \citep{Gopalswamy:2006}. There have been many studies to understand CME initiation with several numerical simulation models. \citet{Chen:2000} showed in their simulation that a CME can be triggered by the localized reconnection between a pre-existing coronal field and a reconnection-favored emerging magnetic flux system with some observational supports \citep[e.g.,][]{Feynman:1995,Jing:2004,Wang:2004}. A flux cancellation model \citep[e.g.,][]{vanBallegooijen:1989,Linker:2001}, in agreement with some observations \citep[e.g.,][]{Martin:1994,Gaizauskas:1997,Martin:1998}, suggested that flux cancellation at the neutral line of a sheared arcade can lead to the formation, destabilization, and eruption of a filament followed by a CME. \citet{Antiochos:1999} proposed a model (so-called breakout model) in which the reconnection of a sheared arcade with an overlying background magnetic field gradually removes a constraint over the sheared arcade so that a CME can occur. A kink instability \citep[e.g.,][]{Hood:1981,Fan:2004,Torok:2005,Cho:2009} and a torus instability \citep[e.g.,][]{Bateman:1978,Titov:1999,Kliem:2006} by an emergence of a twisted flux tube were also considered to explain the initiation of CMEs, e.g., \citet{Fan:2007} performed isothermal MHD simulations of the three-dimensional evolution of coronal magnetic fields as a twisted magnetic flux tube emerges gradually into a pre-existing coronal arcade. In addition, there are some other trigger mechanisms: a shear (or twist) motion of footpoints of magnetic arcades \citep[e.g.,][]{Mikic:1988,Kusano:2004}, slow decay of a background magnetic field \citep[e.g.,][]{Isenberg:1993}, a buoyancy force due to filament mass drainage \citep[e.g.,][]{Low:2001,Zhou:2006}, and Moreton \& EIT waves generated by a remote CME \citep[e.g.,][]{Ballester:2006}. However, a precondition and a trigger mechanism for CME initiation has not been understood clearly yet. More observational studies are needed to determine and/or suggest a most convincing model which can explain the detailed processes of CME occurrence.

After CMEs are initiated at the corona, they depart from the Sun at speeds ranging from $\sim$20 to more than 3000$\,$km$\,$s$^{-1}$ (average speed of 480$\,$km$\,$s$^{-1}$) measured from the $SOHO$/LASCO white-light images during solar cycle 23 \citep{Gopalswamy:2006}. To understand how CME speed is associated with magnetic properties in CME-productive active regions, there have been some studies to find a relationship between CME speed and several magnetic parameters derived from photospheric magnetic fields. \citet{Qiu:2005} studied 13 CME events and found that there is a strong correlation between CME speed and total reconnection flux \citep[see][and references therein]{Forbes:2000} estimated from photospheric magnetic fields with a linear correlation coefficient (CC) of 0.89. \citet{Guo:2007} examined properties of photospheric line-of-sight magnetic fields in 55 active regions before the onset times of 86 CMEs originating from the investigated active regions. They measured four magnetic parameters in the active regions (i.e., tilt angle, total flux, total length of strong-field strong-gradient neutral lines, and effective distance) and found a moderate correlation between CME speed and effective distance with a linear CC of $\sim$0.4.

Magnetic helicity has been recognized as a quantitative and useful measure for global complexity and non-potentiality of a given magnetic field system in terms of twists, kinks, and inter-linkages of magnetic field lines \citep{Berger:1984,Pevtsov:2008,Demoulin:2009}. However, there have been only a few studies of magnetic helicity that is related to solar eruptive phenomena such as filament eruptions and CMEs. By studying a filament in the active region NOAA 8375 on 1998 November 4, \citet{Romano:2003} found that the filament has a helically twisted configuration (shown by a TRACE 171 {\AA} image) and there is a steady magnetic helicity injection of 2$\times$10$^{41}\,$Mx$^2$ from the photosphere of the active region into the corona over a time span of $\sim$28 hours around the times of four eruptive events involved in the filament. From the study of 34 pairs of CMEs and magnetic clouds (MC), \citet{Sung:2009} showed that the square of CME speed is positively correlated with magnetic helicity per unit length in the corresponding MC at 1 AU. \citet{Smyrli:2010} investigated the temporal variation of photospheric magnetic helicity injection in 10 active regions which produced 12 halo CMEs. They found that there is no unique trend of a short-term (several hours) variation of magnetic helicity injection before and after CME occurrence even though a rapid and large change of magnetic helicity injection was observed in some cases. Recently, \citet{Romano:2011} examined a filament eruption in the active region NOAA 9682 and reported that the filament eruption may be caused by the interaction of two magnetic field systems with opposite signs of magnetic helicity.

In this study, long-term variation (a few days) of magnetic helicity injection is investigated for 28 active regions which produced 47 CMEs. The goals of this study are: (1) to find a characteristic variation pattern of helicity injection in relation to a pre-CME condition and a CME energy build-up process and (2) to carry out a correlation study between CME speed and helicity injection rate in the active regions. The rest of this paper is organized as follows: data selection and analysis are explained in Section~\ref{sec2}. In Section~\ref{sec3}, we first describe in detail the observational finding of two CME groups classified by two characteristic evolution patterns of helicity injection and their different CME characteristics (\S~\ref{sec3:sub1}), then the relationship of CME speed with helicity injection rate for the two CME groups is presented (\S~\ref{sec3:sub2}). Finally, we discuss in Section~\ref{sec4} that the two helicity patterns can be interpreted by some numerical simulation models for CME initiation.

\section{Data Selection and Analysis}
\label{sec2}
For a statistical study of magnetic helicity injection in active regions related to CMEs, we selected 28 active regions which produced 47 CMEs (refer to Table~\ref{sec2:tab1} for the detailed information of the CME events). Out of total 47 CMEs, 42 CMEs were adopted from the CME list of \citet{Guo:2007} in which the CMEs are identified with their originating active regions. We carefully checked the identification of the CME source regions by investigating not only the CME position angles with respect to their source regions but also CME-related phenomena in the source regions such as soft X-ray flares, EIT brightenings/dimmings, erupting filaments, or post-flare arcades. The same method for the CME source region identification was applied to add five more CME samples in this study.

By helicity, we refer to the {\it relative} magnetic helicity in the rest of this paper, i.e., the amount of helicity subtracted by that of the corresponding potential field state. The injection rate of helicity $\dot{H}$ through the photospheric surface $S$ of the CME-productive active regions at a specific time $t$ is calculated from the method developed by \citet{Chae:2001}:
\begin{equation}
\label{sec2:equ1}
\dot{H}(t) \simeq \int_{S} G_{A}(\textit{\textbf{x}},t)\,dS,
\end{equation}
where the integrand $G_{A}(\textit{\textbf{x}},t)$ = $-2[\textit{\textbf{v}}_{\mathrm{LCT}}(\textit{\textbf{x}},t) \cdot {\textit{\textbf{A}}}_{p}(\textit{\textbf{x}},t)]B_{n}(\textit{\textbf{x}},t)$ is a good proxy of helicity flux density (i.e., helicity injection per unit area per unit time) at a point $\textit{\textbf{x}}$ on $S$. $G_{A}(\textit{\textbf{x}},t)$ is calculated using MDI magnetogram data with a spatial resolution of 2$\arcsec$ per pixel as follows: (1) $B_{n}(\textit{\textbf{x}},t)$, the magnetic field perpendicular to $S$, is derived from $B_{l}(\textit{\textbf{x}},t)$, the MDI line-of-sight magnetic field, based on the assumption that the transverse component of the magnetic field on $S$ is negligible compared to the longitudinal component (i.e., $B_{n}(\textit{\textbf{x}},t)$ = $B_{l}(\textit{\textbf{x}},t)$/$\cos{\psi(\textit{\textbf{x}})}$ where $\psi(\textit{\textbf{x}})$ is the heliocentric angle of the point \textit{\textbf{x}}). However, to reduce the uncertainty of determining $B_{n}(\textit{\textbf{x}},t)$, we calculate $G_{A}(\textit{\textbf{x}},t)$ only if $\textit{\textbf{x}}$ is located within 0.6 of a solar radius from the apparent disk center, (2) $\textit{\textbf{v}}_{\mathrm{LCT}}(\textit{\textbf{x}},t)$, the velocity of the apparent horizontal motion of field lines on $S$, is determined by the technique of local correlation tracking (LCT) using two consecutive 96-min MDI magnetogram data, and (3) $\textit{\textbf{A}}_{p}(\textit{\textbf{x}},t)$ is a specific vector potential satisfying:
\begin{equation}
\label{sec2:equ2}
{\hat{\textit{\textbf{n}}}} \cdot \nabla \times {\textit{\textbf{A}}}_{p}(\textit{\textbf{x}},t) = B_{n}(\textit{\textbf{x}},t),
\end{equation}
\begin{equation}
\label{sec2:equ3}
\nabla \cdot {\textit{\textbf{A}}}_{p}(\textit{\textbf{x}},t) = 0,
\end{equation}
\begin{equation}
\label{sec2:equ4}
{\textit{\textbf{A}}}_{p}(\textit{\textbf{x}},t) \cdot {\hat{\textit{\textbf{n}}}} = 0.
\end{equation}
$\textit{\textbf{A}}_{p}(\textit{\textbf{x}},t)$ is calculated by using the fast Fourier transform (FFT) method.

\begin{figure}
\begin{center}\epsscale{1}
\plotone{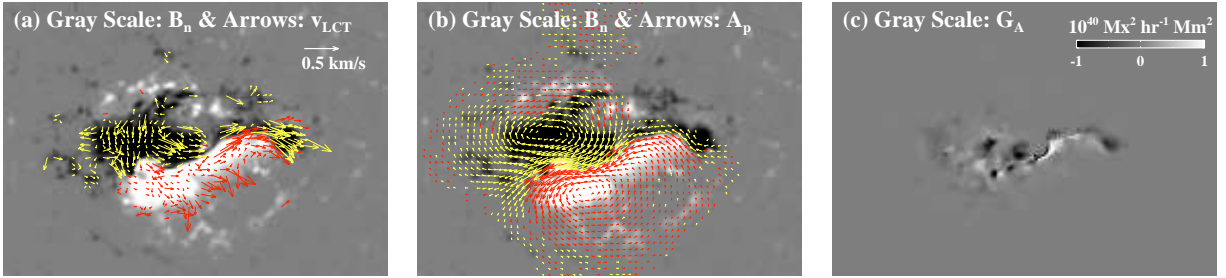}
\caption{Solar active region NOAA 10720 at 14:23 UT on 2005 January 16. Panels (a) and (b) show $\textit{\textbf{v}}_{\mathrm{LCT}}$ (red/yellow arrows on positive/negative $B_{n}$) and $\textit{\textbf{A}}_{p}$ (red/yellow arrows on positive/negative $B_{n}$) superposed on the grayscale image of $B_{n}$ derived from the MDI line-of-sight magnetogram, respectively. In Panel (c), $G_{A}$ map is presented in grayscale. Note that the saturation level of $|G_{A}|$ is set as 1$\times$10$^{40}\,$Mx$^{2}\,$hr$^{-1}\,$Mm$^{-2}$ for purpose of display visibility.}
\label{sec2:fig1}
\end{center}
\end{figure}

Figure~\ref{sec2:fig1}{a} and~\ref{sec2:fig1}{b} present $\textit{\textbf{v}}_{\mathrm{LCT}}$ and $\textit{\textbf{A}}_{p}$ of AR 10720, respectively, which are plotted as arrows on the grayscale image of $B_{n}$ at 14:23 UT on 2005 January 16. The grayscale $G_{A}$ map is also shown in Figure~\ref{sec2:fig1}{c}: positive value (i.e., helicity flux density of right-handed sign) of $G_{A}$ is displayed as white tone, while negative value (i.e., helicity flux density of left-handed sign) of $G_{A}$ is displayed as black tones. See \citet{Chae:2005} for the details of the $G_{A}$ calculation. Many previous studies have taken account of $G_{A}$ for calculating helicity injection in solar active regions \citep[e.g.,][]{Moon:2002a,Moon:2002b,Nindos:2003,Chae:2004,Romano:2005,Jeong:2007,LaBonte:2007,Lim:2007,Park:2008,Park:2010}.
Note that although a better proxy of helicity flux density $G_{\theta}$ was proposed by \citet{Pariat:2005}, however, by the comparison between these two methods ($G_{A}$ and $G_{\theta}$) of $\dot{H}$ calculation, \citet{Chae:2007} found that their discrepancy is typically less than 10$\,\%$. \citet{Park:2010} found that the relative standard deviation of $\dot{H}$ calculated from the same method as used in this study is around 5$\,\%$. We therefore think that the uncertainty of $G_{A}$ does not significantly affect the helicity calculation and conclusion of this study.

After $\dot{H}$ is determined as a function of time, the accumulated amount of helicity injection $\Delta H$ in the CME-productive active regions is determined by integrating $\dot{H}$ with respect to time:
\begin{equation}
\label{sec2:equ5}
\Delta H(t) = \int_{t_0}^{t} \dot{H}(t')\,dt',
\end{equation}
where $t_0$ is the start time of each MDI magnetogram data set of the active region under investigation. $t_0$ is set as the time when the active region appears or rotates to a position within 0.6 of a solar radius from the apparent disk center. As a reference parameter, we also derive the total unsigned magnetic flux $\Phi$ at $S$:
\begin{equation}
\label{sec2:equ6}
\Phi(t) = \int_{S} |B_{n}(\textit{\textbf{x}},t)|\,dS.
\end{equation}

\section{Results}
\label{sec3}

\subsection{Two Characteristic Patterns of Helicity Injection before CMEs}
\label{sec3:sub1}
We have investigated helicity injection through the photospheric surface of 28 solar active regions over a span of several days around the times of 47 CMEs which originated from the active regions. As a result, it was found that all the CMEs under investigation are always preceded by a significant helicity injection of 10$^{42}$--10$^{44}\,$Mx$^2$ over a long period ($\sim$0.5--4.5 days) into the active-region corona through the photosphere. Note that there have been previous reports of a large amount of helicity injection in active regions for $\sim$0.5--4.5 days before the occurrence of major flares \citep[e.g.,][]{LaBonte:2007,Park:2008,Park:2010}, a filament eruption \citep[e.g.,][]{Romano:2003}, and halo CMEs \citep[e.g.,][]{Smyrli:2010}. Furthermore, we found that there are two characteristic patterns in the long-term (a few days) variation of helicity injection in the CME-productive active regions which appeared for a few days before the CME occurrence: (1) a monotonically increasing pattern with one sign of helicity (Group A; 30 CMEs in 23 active regions) and (2) a pattern of significant helicity injection followed by its sign reversal (Group B; 17 CMEs in 5 active regions).

\begin{figure}
\begin{center}\epsscale{0.65}
\plotone{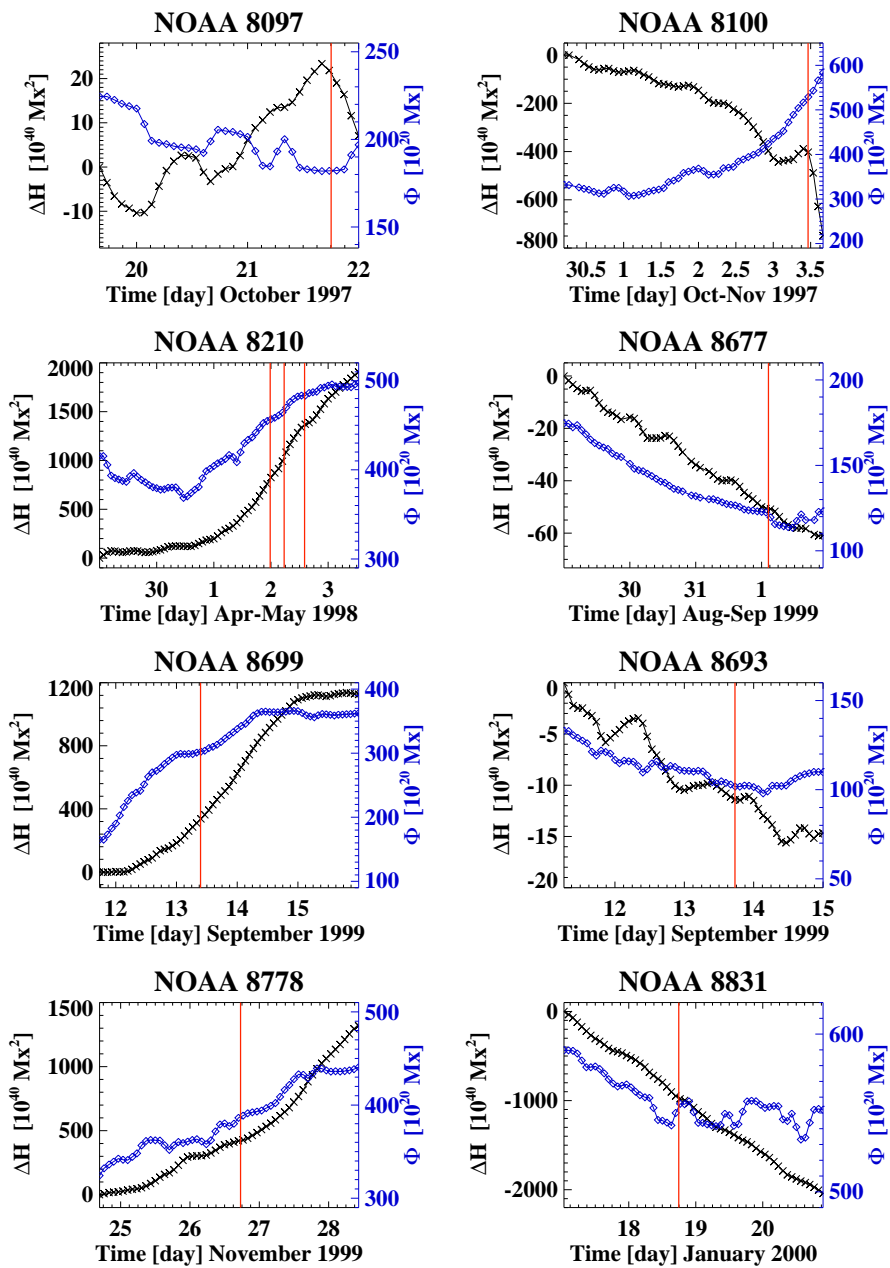}
\caption{Time variation of helicity injection $\Delta H$ (black crosses) and total unsigned magnetic flux $\Phi$ (blue diamonds) for 8 active regions in Group A. The active regions in Group A show a monotonically increasing pattern of helicity for a few days. In each panel, the vertical red lines indicate the times when the CMEs originating from the 8 active regions first appeared in the LASCO/C2 FOV.}
\label{sec3:sub1:fig1}
\end{center}
\end{figure}

\begin{figure}
\begin{center}\epsscale{0.65}
\plotone{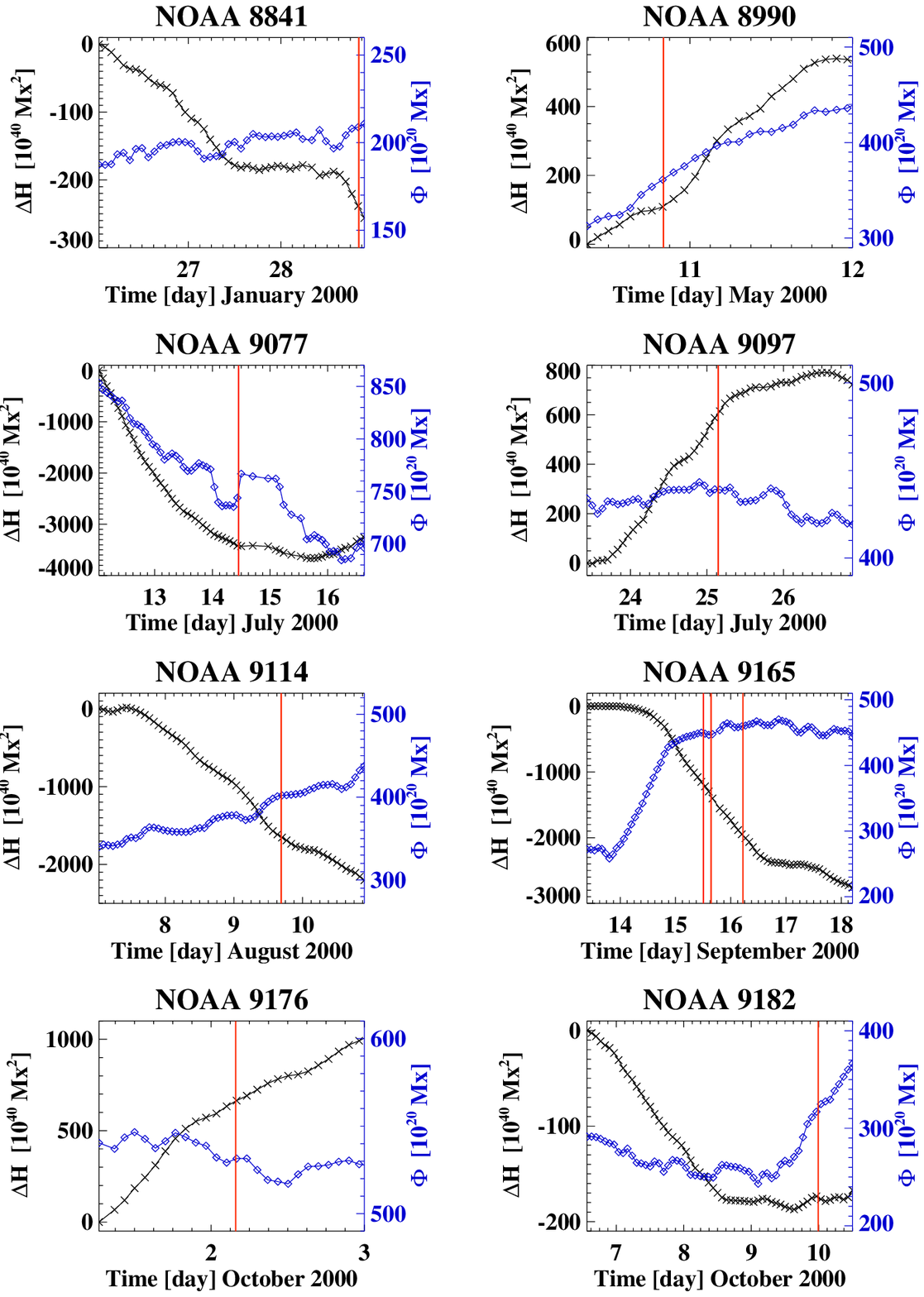}
\caption{Same as in Figure~\ref{sec3:sub1:fig1}, but for additional 8 active regions in Group A.}
\label{sec3:sub1:fig2}
\end{center}
\end{figure}

\begin{figure}
\begin{center}\epsscale{0.65}
\plotone{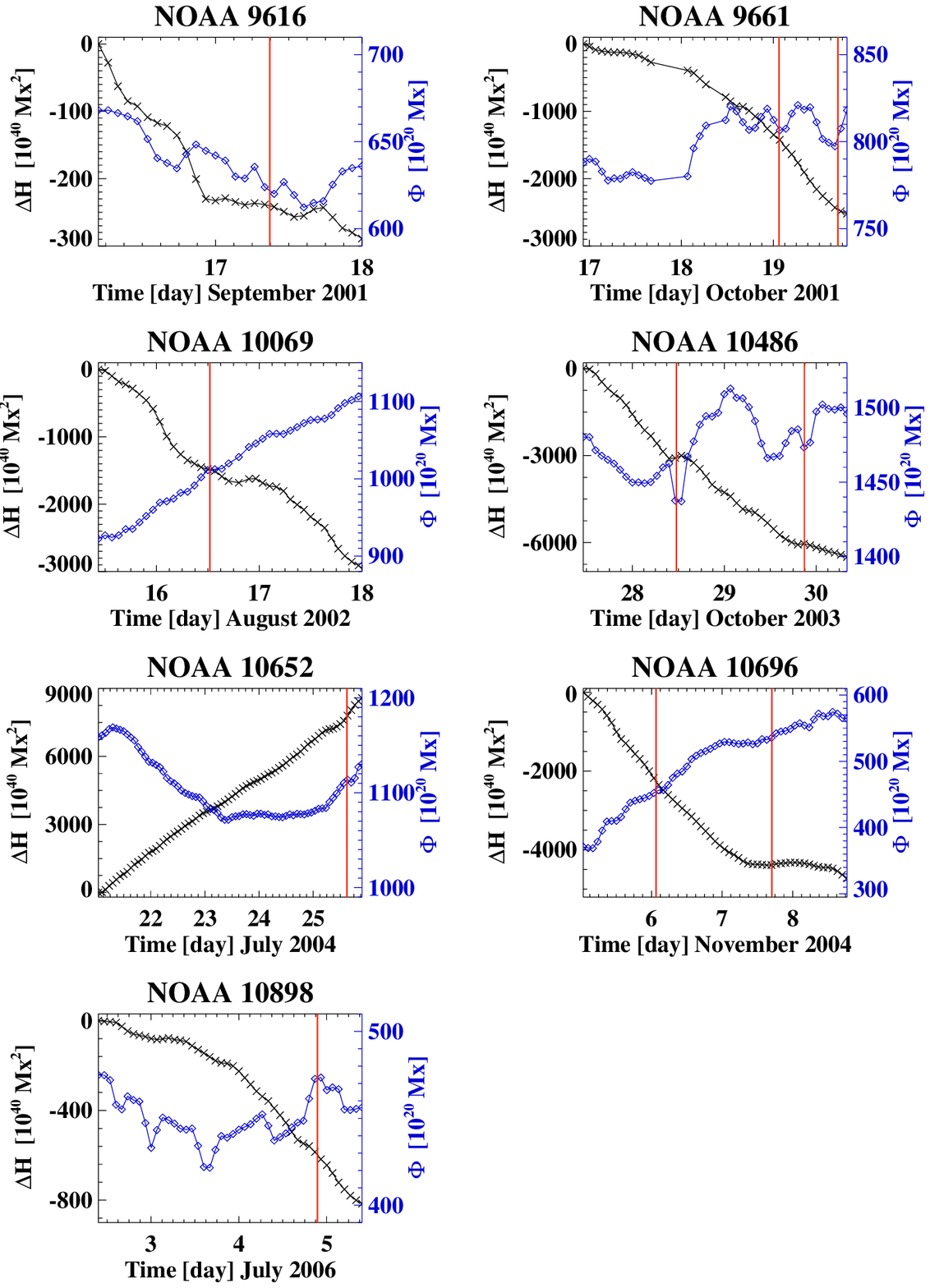}
\caption{Same as in Figure~\ref{sec3:sub1:fig1}, but for additional 7 active regions in Group A.}
\label{sec3:sub1:fig3}
\end{center}
\end{figure}

\begin{figure}
\begin{center}\epsscale{0.65}
\plotone{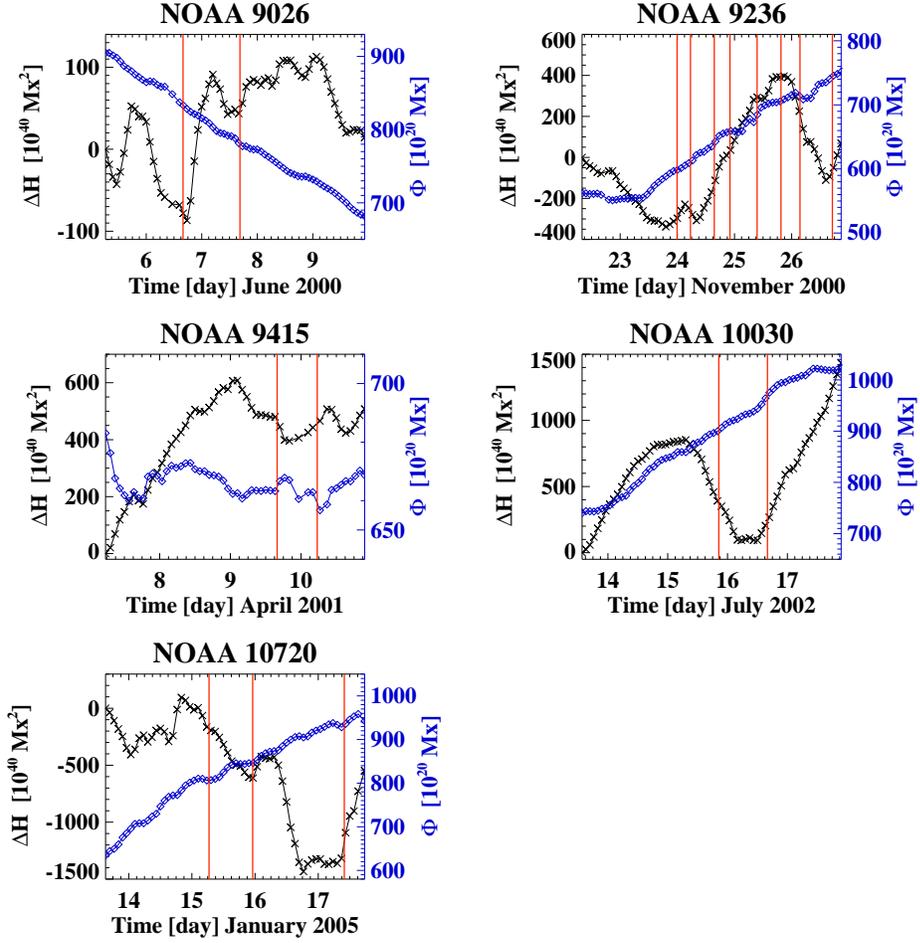}
\caption{Same as in Figure~\ref{sec3:sub1:fig1}, but for 5 active regions in Group B indicating a pattern of significant helicity injection followed by its sign reversal. A total of 17 CMEs occurred from the 5 active regions during the period when the helicity injection rate in the active regions started to reverse its sign.}
\label{sec3:sub1:fig4}
\end{center}
\end{figure}

Figures~\ref{sec3:sub1:fig1},~\ref{sec3:sub1:fig2}, and~\ref{sec3:sub1:fig3} present the time profiles of $\Delta H$ calculated for the 23 active regions in Group A. As shown in the time profiles, each of the 30 CMEs in Group A was preceded by a large injection of one sign of helicity. More specifically, the absolute value of $\Delta H$ significantly increased at a nearly constant rate, (0.3--110)$\times$10$^{40}\,$Mx$^{2}\,$hr$^{-1}$, over about 0.5--4.5 days and then 26 out of the 30 CMEs occurred during the monotonically increasing phase. The other 4 CMEs in Group A occurred when the helicity increasing rate becomes much smaller (almost zero) after the monotonically increasing phase, and they are marked with the superscript `$\dag$' in Table~\ref{sec2:tab1}. On the other hands, the time variations of $\Delta H$ for the 5 active regions in Group B are shown in Figure~\ref{sec3:sub1:fig4}. The source regions producing 17 CMEs in Group B presented a sign change of $\dot{H}$ before the CMEs: i.e., a noticeable increase in helicity of one sign was present for a while (6--45 hours) and then the CMEs occurred while relatively more helicity with the opposite sign was being injected in their source regions for the next few days. The main difference between Group A and Group B in the temporal variation of helicity injection is that the helicity sign reversal phase appears in Group B, but not in Group A. Note that we made the criterion for the helicity sign reversal shown in Group B as $\Delta H$ changes greater than 10$^{42}\,$Mx$^2$ before and after the sign reversal phase starts. In the case that a sign change of $\dot{H}$ appeared but the criterion was not satisfied, we considered the variation of $\Delta H$ as fluctuation during the monotonically increasing phase of helicity injection shown in the active regions NOAA 8097, 8100, and 8693 of Group A.

We have also examined whether or not there is a specific trend in the temporal variation of $\Phi$ in the CME source active regions during the period of the two characteristic helicity patterns shown in Group A and Group B. For doing that, from the time profiles of $\Phi$ shown in Figures~\ref{sec3:sub1:fig1},~\ref{sec3:sub1:fig2},~\ref{sec3:sub1:fig3}, and~\ref{sec3:sub1:fig4}, we calculated $\dot{\Phi}_{avg}$ that is the average change rate of the total unsigned magnetic flux during the time period $\Delta\tau$ between $t_{0}$ and the CME occurrence time $t_{1}$ in the active region under investigation: 
\begin{equation}
\label{sec3:sub1:equ1}
\dot{\Phi}_{avg} = \frac{[\Phi(t_{1})-\Phi(t_{0})]}{\Delta\tau}.
\end{equation}
In the case of the 30 CMEs in Group A, 19 CMEs (63$\,\%$) were preceded by an increase in $\Phi$; i.e., $\dot{\Phi}_{avg}$ $>$ 0. However, the other 11 CMEs (37$\,\%$) were preceded by a decrease in $\Phi$; i.e., $\dot{\Phi}_{avg}$ $<$ 0. For the 17 CMEs in Group B, 13 CMEs (76$\,\%$) occurred after an increase in $\Phi$, while 4 CMEs (24$\,\%$) after a decrease in $\Phi$. Refer to Tables~\ref{sec3:sub2:tab1} and~\ref{sec3:sub2:tab2} for the detailed values of $\Phi(t_{1})$, $\dot{\Phi}_{avg}$, and $\Delta\tau$ of each event in Group A and Group B. We found that an increasing trend of $\Phi$ before the occurrence of the CMEs is more common than a decreasing trend and this is more apparently shown in Group B than Group A. 

Now it would be interesting to know whether CME characteristics are different according to the two different patterns of helicity injection in the CME source regions. We therefore investigated a difference in the speed, acceleration, and occurrence trend of the CMEs between Group A and Group B. The CME speed and acceleration were adopted from the LASCO CME catalog (\url{http://cdaw.gsfc.nasa.gov/CME_list}). The linear speed in the catalog was taken as the representative CME speed in this study, which was determined from linear fit (also known as constant speed fit) to the height-time measurement of each of the CMEs  within the LASCO C2 and C3 field of view (1.5--30 solar radii from the solar surface). The CME acceleration was also estimated from the height-time measurement using second-order polynomial fit. Note that there is a large uncertainty in some cases of the CME acceleration measurement with just three points in the height-time plot and they are marked with the superscript `$\ast$' in Table~\ref{sec2:tab1}.

We found that the CMEs in Group A have the speed range of 250--2460$\,$km$\,$s$^{-1}$ with the average(median) speed of 870(700)$\,$km$\,$s$^{-1}$ while those in Group B have the speed range of 500--2860$\,$km$\,$s$^{-1}$ with the average(median) speed of 1330(1150) km$\,$s$^{-1}$. The average CME speed of Group B is about 450$\,$km$\,$s$^{-1}$ faster than that of Group A. There is also a significant difference in the CME acceleration between Group A and Group B: the average acceleration is -24.4$\,$m$\,$s$^{-2}$ and -6.3$\,$m$\,$s$^{-2}$, respectively for Group A and Group B. These facts indicate that the CMEs in Group A have a relatively slow speed and rapid deceleration compared to those in Group B. Furthermore, the CMEs in Group A mainly consist of single events, while those in Group B tend to be successive events; i.e., 18 out of the total 23 active regions in Group A produced only one CME event during the helicity measurement period, while each of the total 5 active regions in Group B generated at least two CMEs. We examined soft X-ray characteristics of the CME-related flares in Group A and Group B using the 1--8 {\AA} $GOES$ data. Both groups have a similar average flare duration, but they show a difference in total integrated flux $F$ and flare impulsiveness $I$ \citep[i.e., peak flux/flare rise time; refer to][]{Pearson:1989}: the average $F$ (10$^{-5}\,$W$\,$m$^{-2}\,$min$^{-1}$ ) is 1.5 and 2.3 and the average $I$ (10$^{-5}\,$W$\,$m$^{-2}\,$min$^{-1}$ ) is 0.7 and 1.1, respectively for Group A and Group B. This indicates that the CME-related flares in Group B are relatively more energetic and impulsive than those in Group A. In Table~\ref{sec3:sub1:tab1}, the differences between Group A and Group B are summarized.

\subsection{Correlation bewteen CME Speed and Helicity Injection Rate}
\label{sec3:sub2}
From the time profile of $\dot{H}$, a helicity parameter is defined to investigate its relationship with CME speed. We use the average helicity injection rate $\dot{H}_{avg}$ which indicates the average amount of injected helicity per unit time into $S$ during the time period $\Delta\tau$ between $t_{0}$ and $t_{1}$:
\begin{equation}
\label{sec3:sub2:equ7}
\dot{H}_{avg} = \frac{\sum_{t_{0}}^{t_{1}}\dot{H}(t)}{N},
\end{equation}
where $N$ is the total number of data points during $\Delta\tau$.
Note that as a reference parameter, we consider $\Phi(t_{1})$, i.e., the total unsigned magnetic flux at $t_{1}$:
\begin{equation}
\Phi(t_{1}) = \int_{S} |B_{n}(\textit{\textbf{x}},t_{1})| \, dS.
\end{equation}

\begin{figure}
\begin{center}\epsscale{0.9}
\plotone{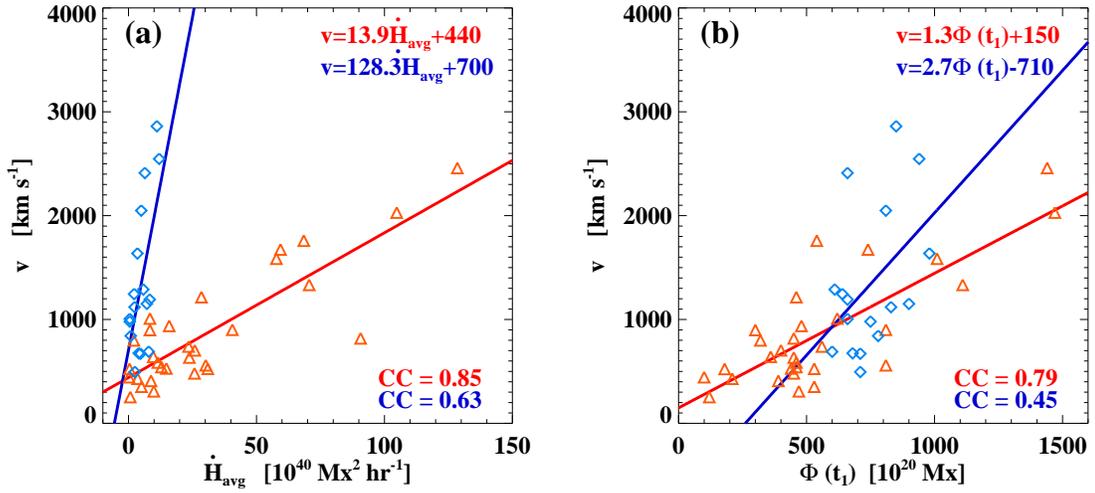}
\caption{The speed $v$ of 47 CMEs (30 CMEs in Group A marked by red triangles and 17 CMEs in Group B marked by blue diamonds) is plotted against the two magnetic parameters of (a) the average helicity injection rate $\dot{H}_{avg}$ and (b) the total unsigned magnetic flux $\Phi(t_{1})$. In each panel, the red and blue solid lines indicate the least-square linear fits to the data points of Group A and Group B, respectively, and the slope and intercept of the fitted lines are specified with the linear correlation coefficient (CC). The correlation between $v$ and $\dot{H}_{avg}$ is very strong (CC=0.85) for Group A and moderate (CC=0.63) for Group B. See Table~\ref{sec3:sub2:tab1} for the detailed information of the 47 CMEs.}
\label{sec3:sub2:fig1}
\end{center}
\end{figure}

\begin{figure}
\begin{center}\epsscale{0.9}
\plotone{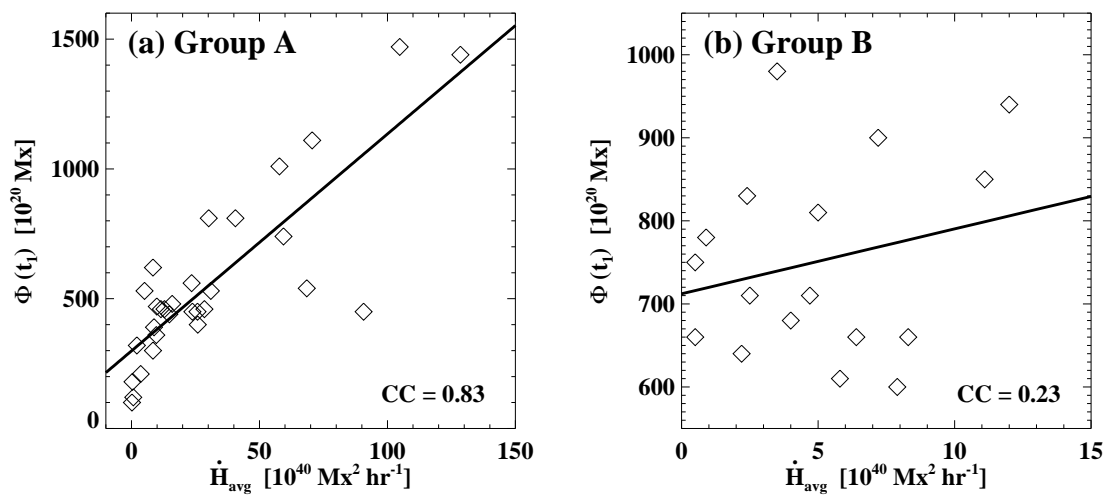}
\caption{The total unsigned flux $\Phi(t_{1})$ vs. the average helicity injection rate $\dot{H}_{avg}$ for (a) 30 CMEs in Group A and (b) 17 CMEs in Group B. The least-square linear fit (solid line) and the linear correlation coefficient (CC) are presented in each panel. A very high correlation between $\dot{H}_{avg}$ and $\Phi(t_{1})$ is found for Group A (CC=0.83), but a poor correlation for Group B (CC=0.23).}
\label{sec3:sub2:fig2}
\end{center}
\end{figure}

The two parameters $\dot{H}_{avg}$ and $\Phi(t_{1})$ were calculated for the source regions producing the 47 CMEs under investigation, and they were compared with the CME speed $v$ as shown in Figure~\ref{sec3:sub2:fig1}. The red and blue solid lines indicate the least-squares linear fits to the data points of Group A (red triangles) and Group B (blue diamonds) in each panel, respectively. The slope, intercept, and CC of the linear fits are also given in each panel. For Group A, we found that $v$ has a very strong correlation (CC=0.85) with $\dot{H}_{avg}$, as well as a good correlation (CC=0.79) with $\Phi(t_{1})$. In case of Group B, $v$ has a moderate correlation (CC=0.63) with $\dot{H}_{avg}$, while its correlation with $\Phi(t_{1})$ is weak (CC=0.45). This is because there is a high correlation (CC=0.83) between $\dot{H}_{avg}$ and $\Phi(t_{1})$ for Group A (see Figure~\ref{sec3:sub2:fig2}{a}) as previously reported that there is a good correlation between helicity injection and total unsigned magnetic flux in active regions \citep{Jeong:2007,Park:2010}, but a poor correlation (CC=0.23) for Group B (as shown in Figure~\ref{sec3:sub2:fig2}{b}). Note that the correlations between $v$ and $\dot{H}_{avg}$ for both Group A and Group B are statistically highly significant with $>$ 99$\,\%$ confidence level (from a two-tailed Student's t-test). In addition, we found that the linear fits to the data points of $v$ (km$\,$s$^{-1}$) vs. $\dot{H}_{avg}$ (10$^{40}\,$Mx$^{2}\,$hr$^{-1}$) are significantly different between Group A and Group B: i.e., $v$=14$\dot{H}_{avg}$+440 for Group A and $v$=128$\dot{H}_{avg}$+700 for Group B. The slope of the linear fitted line for Group B is about nine times greater than that for Group A, and the intercept of the fitted line for Group B is $\sim$250 km/s greater than that for Group A.

We now have a question why there are considerable correlations between the helicity parameter and the CME speed with a tendency that the lager injection of helicity a CME-productive active region achieves, the faster CME it produces. One possible answer for the question is as follows: the speed of CMEs depends on how much energy CMEs have, and the CME kinetic energy is supposed to be originated from free magnetic energy which is the energy deviation of the coronal magnetic field from its potential state. Assuming linear force-free magnetic fields, \citet{Georgoulis:2007} showed that there is linear dependence between free magnetic energy and relative magnetic helicity from the total magnetic energy formula given by \citet{Berger:1988}. In the study of \citet{Regnier:2007}, it was also found that there is a correlation between free magnetic energy and relative magnetic helicity estimated from reconstructed nonlinear force-free magnetic fields in four active regions even though the number of the active region samples is too small. Therefore, it might be reasonable to consider that helicity injection through the photosphere in a CME-productive active region is closely related to free magnetic energy build-up in the active-region corona for CME kinetic energy.

\section{Discussion and Conclusion}
\label{sec4}
47 CME events are examined to find a characteristic evolution pattern of helicity injection during a few days in their 28 source active regions. The main findings in this study are as follows: (1) there is always a significant helicity injection of 10$^{42}$--10$^{44}\,$Mx$^2$ through the active-region photosphere over a few days before the CMEs; (2) the CMEs under investigation are categorized into two different groups by two different helicity patterns which appeared for $\sim$0.5--4.5 before their occurrence. 30 CMEs in Group A occurred in 23 active regions in which monotonically increasing pattern with one sign of helicity injection is presented, whereas 5 active regions producing 17 CMEs in Group B show a pattern of significant helicity injection followed by its sign reversal; and (3) the correlation between CME speed $v$ and average helicity injection rate $\dot{H}_{avg}$ in the CME-productive active regions is significant (CC=0.85) for Group A and moderate (CC=0.63) for Group B.

In addition, the two CME groups classified by the two characteristic evolution patterns of helicity injection show different characteristics in kinematics, occurrence rate in a single active region, CME-related flare properties, and magnetic properties of their source regions as follows: (1) the average CME speed of Group B (1330$\,$km$\,$s$^{-1}$) is much faster than that of Group A (870$\,$km$\,$s$^{-1}$) and the CMEs in Group B (-6.3$\,$m$\,$s$^{-2}$) have a relatively slow deceleration compared to those in Group A (-24.4$\,$m$\,$s$^{-2}$); (2) a linear fit of $v$ vs. $\dot{H}_{avg}$ is quite different between Group A and Group B; (3) the CMEs in Group A tend to be single events, while those in Group B are inclined to be successive events; (4) soft X-ray flares related to the CMEs in Group B are relatively more energetic and impulsive than those in Group A; and (5) an increasing trend of magnetic flux in the CME source regions is more commonly shown in Group B than Group A.

These differences may suggest that there are different pre-CME conditions for the two groups related to the two characteristic helicity patterns. We therefore try to understand how the two helicity patterns are involve with a precondition for CME initiation by comparing them with temporal variations of helicity injection which are calculated or expected from some numerical CME models. First, emerging twisted flux rope models \citep[e.g.,][]{Titov:1999,Fan:2004,Torok:2005,Fan:2007,Isenberg:2007,Fan:2010} can be considered to explain the characteristic variation pattern of $\Delta H$ shown in Group A (i.e., the monotonically increasing pattern with one sign of helicity). \citet{Fan:2004} showed in their numerical MHD simulation that as a twisted flux tube emerges gradually into a pre-existing coronal arcade during a pre-CME phase, the flux tube's helicity is continuously transported through the photosphere boundary into the corona. So we can verify that the time variation of photospheric helicity injection in the simulation is very similar to that in Group A \citep[refer to Figure 3c in][]{Fan:2004}. In addition, the numerical simulations of \citet{Fan:2007} and \citet{Fan:2010} indicated that (1) an emerging flux tube can settle into a phase of a steady quasi-static rise after the emergence of the flux tube is slowed down or stopped and (2) the quasi-static rising phase can be sustained for several hours until it erupts as a CME. We therefore conjecture that the phase of the nearly constant $\Delta H$ which appeared for several hours before the onset time of the 4 CMEs in Group A may be related to the quasi-static stage. Second, some CME models \citep[e.g.,][]{Mikic:1988,Wolfson:1992,Antiochos:1999} which contain pre-CME dynamics associated with a steady shear (or twist) motion of magnetic field line footpoints on the photosphere can also explain the monotonically increasing pattern in Group A. In this case, magnetic flux emergence at the photosphere is not necessary to inject helicity, but helicity can be continuously injected via the steady photospheric shear (or twist) motion.

On the other hand, we consider reconnection-favored emerging flux models \citep[e.g.,][]{Chen:2000,Archontis:2007,Archontis:2008} to explain the characteristic variation pattern of $\Delta H$ shown in Group B (i.e., the helicity sign reversal pattern). The numerical MHD simulation of \citet{Chen:2000} showed that a CME can be initiated by the localized reconnection between a pre-existing coronal field and a reconnection-favored emerging flux. And it is evident that this reconnection-favored flux can provide helicity injection of the opposite sign into an existing helicity system: hence, the phase of the opposite-signed helicity injection can be produced by a steady and significant emergence of the reconnection-favored flux. In addition, as simulated by \citet{Kusano:2003}, a relatively strong shear motion of underlying field footpoints in the reverse direction compared to that of overlying field footpoints can explain the helicity sign reversal phase in Group B. Note also that there are some observational studies on spatial distributions of magnetic and electric current helicities in eruptive solar active regions, which may suggest a source region related to the helicity sign reversal. E.g., from the study of fractional electric current helicity $h_{c}$ (i.e., $B_{z} \cdot (\nabla \times \textit{\textbf{B}})_{z}$) in 9 CME-associated active regions, \citet{Wang:2004} reported that a key emerging flux region or a moving magnetic feature near the main sunspot in the active regions brings up $h_{c}$ with a sign opposite to the dominant sign of the main sunspot with a duration of a few days. \citet{Romano:2011} studied a filament eruption in the active region NOAA 9682 and found that positive magnetic helicity was dominantly injected in the entire active region, while negative helicity was injected in local regions where the filament footpoints were located.

In conclusion, we found that the two CME groups classified by the two characteristic helicity patterns have different pre-CME conditions and different CME characteristics. In addition, by comparing the observational helicity patterns with the expected patterns from some of numerical CME models, we presume that: the pre-CME condition of Group A is associated with an emergence of a twisted flux rope or a steady shear (or twist) motion of magnetic field line footpoints, while the pre-CME condition of Group B is involved with a reconnection-favored emerging flux or a reverse shear motion of magnetic field line footpoints. However, there are still some unanswered questions such as why there are significant differences between Group A and Group B in terms of CME kinematics (e.g., speed and acceleration) and why there is a fairly good correlation between $\dot{H}_{avg}$ and $\Phi(t_{1})$ for the source regions in Group A, but a weak correlation for those in Group B. This helicity study should therefore be carefully checked with aspects shown in several other CME numerical simulations to further understand the physics underlying CME triggering mechanism and dynamics. And a future study will focus on the spatial distribution of helicity flux density and the evolution of detailed magnetic field structures in the investigated active regions to better understand CME initiation. 

Finally, we suggest that these characteristic variation patterns and helicity injection rate in CME-productive active regions can be used for the improvement of CME forecasting: (1) an early warning sign of CME occurrence could be given by the presence of a phase of monotonically increasing helicity as it is found that all the CMEs under investigation occur after significant helicity injection; (2) a warning sign for imminent CME occurrence could be also made when helicity injection rate becomes very slow or the opposite sign of helicity starts to be injected after the significant helicity injection in active regions; and (3) the speed of future CME event can be estimated by the statistical study of the correlation between the CME speed and the average helicity injection rate in the active regions. For doing this, more observational studies are therefore being carried out to check whether the two characteristic helicity pattern are shown in other CME-productive active regions and to improve the correlation between the CME speed and the helcity injection rate.

\acknowledgments The authors thank the $SOHO$/MDI team for the 96-minute full-disk photospheric magnetogram data and the $SOHO$/LASCO team for the CME catalog which is generated and maintained at the CDAW Data Center by NASA and The Catholic University of America in cooperation with the Naval Research Laboratory. SOHO is a project of international cooperation between ESA and NASA. We have made use of NASA's Astrophysics Data System Abstract Service. This work has been support by the ``Development of Korea Space Weather Prediction Center" project of KASI and the KASI basic research fund. RL and HW were supported by NSF grants AGS-0839216 and AGS-0849453 and NASA grants NNX08AJ23G and NNX11AC05G.

\clearpage


\begin{deluxetable}{ccccccccccc}
\tablecolumns{1}
\tabletypesize{\scriptsize}
\tablewidth{0pt}
\tablecaption{List of 47 Selected CME Events \label{sec2:tab1}}
\tablehead{\multirow{3}{*}{No.} & \multicolumn{4}{c}{CME} & \multicolumn{2}{c}{Related Flare} & \multicolumn{2}{c}{Source Region} & \multirow{3}{*}{Group\tablenotemark{f}} \\
& \multirow{2}{*}{$t_a$\tablenotemark{a}} & \colhead{PA/AW\tablenotemark{b}} & \colhead{$v$\tablenotemark{c}} & \colhead{$a$\tablenotemark{d}} &
\multirow{2}{*}{$t_s$}\tablenotemark{e} & GOES & NOAA & Sunspot & \\
& & (deg) & (km$\,$s$^{-1}$) & (m$\,$s$^{-2}$) & & Class & No. & Classification & }
\startdata %
1  &  1997-10-21 18:03 & Halo/360 & 523  & -2.9   & 17:00        & C3.3 & 8097  & $\beta$ & A \\
2  &  1997-11-03 11:11 & 232/122  & 352  & -1.5   & 09:03        & M1.4 & 8100  & $\beta\gamma\delta$ & A$^{\dag}$ \\
3  &  1998-05-01 23:40 & Halo/360 & 585  &  8.0   & 21:40        & C2.6 & 8210  & $\gamma\delta$ & A \\
4  &  1998-05-02 05:31 & Halo/360 & 542  & -1.4   & 04:48        & C5.4 & 8210  & $\beta\gamma\delta$ & A \\
5  &  1998-05-02 14:06 & Halo/360 & 938  & -28.8  & 13:31        & X1.1 & 8210  & $\beta\gamma\delta$ & A \\
6  &  1999-09-01 02:30 & 188/283  & 253  & -1.2*  & 00:27        & C2.7 & 8677  & $\beta$ & A \\
7  &  1999-09-13 09:30 & 0/182    & 898  & -23    & 08:05        & C4.9 & 8699  & $\beta$ & A \\
8  &  1999-09-13 17:31 & 109/184  & 444  & -8.7*  & 16:30        & C2.6 & 8693  & $\beta$ & A \\
9  &  1999-11-26 17:30 & 228/145  & 409  & 6.0    & 17:40        & C2.3 & 8778  & $\beta$ & A \\
10  & 2000-01-18 17:54 & Halo/360 & 739  & -7.1   & 17:07        & M3.9 & 8831  & $\beta$ & A \\
11  & 2000-01-28 20:12 & 70/20    & 429  & -2.8   & 19:45        & C4.7 & 8841  & $\beta$ & A \\
12  & 2000-05-10 20:06 & 83/205   & 641  & -15.5  & 19:26        & C8.7 & 8990  & $\beta$ & A \\
13  & 2000-06-06 15:54 & Halo/360 & 1119 & 1.5    & 14:58        & X2.3 & 9026  & $\beta\gamma\delta$ & B \\
14  & 2000-06-07 16:30 & Halo/360 & 842  & 59.8*  & 15:34        & X1.2 & 9026  & $\beta\gamma\delta$ & B \\
15  & 2000-07-14 10:54 & Halo/360 & 1674 & -96.1* & 10:03        & X5.7 & 9077  & $\beta\gamma\delta$ & A \\
16  & 2000-07-25 03:30 & Halo/360 & 528  & -5.8   & 02:43        & M8.0 & 9097  & $\beta\gamma$ & A \\
17  & 2000-08-09 16:30 & Halo/360 & 702  & 2.8    & 15:19        & C2.3 & 9114  & $\beta\gamma$ & A \\
18  & 2000-09-15 12:06 & 249/235  & 633  & -64.0* & 10:51        & C9.5 & 9165  & $\beta\delta$ & A \\
19  & 2000-09-15 15:26 & 217/210  & 481  & -10.4* & 14:29        & M2.0 & 9165  & $\beta\delta$ & A \\
20  & 2000-09-16 05:18 & Halo/360 & 1215 & -12.3  & 04:06        & M5.9 & 9165  & $\beta\gamma$ & A \\
21  & 2000-10-02 03:50 & Halo/360 & 525  & -4.9   & 02:48        & C4.1 & 9176  & $\beta\gamma$ & A \\
22  & 2000-10-09 23:50 & Halo/360 & 798  & -9.8   & 23:19        & C6.7 & 9182  & $\beta$ & A$^{\dag}$ \\
23  & 2000-11-23 23:54 & 336/157  & 690  & 1.8    & 23:18        & M1.0 & 9236  & $\beta$ & B \\
24  & 2000-11-24 05:30 & Halo/360 & 1289 & 2.1    & 04:55        & X2.0 & 9236  & $\beta\gamma$ & B \\
25  & 2000-11-24 15:30 & Halo/360 & 1245 & -3.3   & 14:51        & X2.3 & 9236  & $\beta\gamma$ & B \\
26  & 2000-11-24 22:06 & Halo/360 & 1005 & -0.8   & 21:43        & X1.8 & 9236  & $\beta\gamma$ & B \\
27  & 2000-11-25 09:30 & Halo/360 & 675  & -4.7   & 09:06        & M3.5 & 9236  & $\beta\gamma$ & B \\
28  & 2000-11-25 19:31 & Halo/360 & 671  & -10.8  & 18:33        & X1.9 & 9236  & $\beta\gamma$ & B \\
29  & 2000-11-26 03:30 & 259/118  & 495  & -22.9* & 02:47        & M2.2 & 9236  & $\beta\gamma$ & B \\
30  & 2000-11-26 17:06 & Halo/360 & 980  & 5.8    & 16:34        & X4.0 & 9236  & $\beta\gamma$ & B \\
31  & 2001-04-09 15:54 & Halo/360 & 1192 & 1.3    & 15:20        & M7.9 & 9415  & $\beta\gamma\delta$ & B \\
32  & 2001-04-10 05:30 & Halo/360 & 2411 & 211.6* & 05:06        & X2.3 & 9415  & $\beta\gamma\delta$ & B \\
33  & 2001-09-17 08:54 & 198/166  & 1009 & -14.5  & 08:18        & M1.5 & 9616  & $\beta\gamma$ & A$^{\dag}$ \\
34  & 2001-10-19 01:27 & Halo/360 & 558  & -25.6  & 00:47        & X1.6 & 9661  & $\beta\gamma\delta$ & A \\
35  & 2001-10-19 16:50 & Halo/360 & 901  & -0.7   & 16:13        & X1.6 & 9661  & $\beta\gamma\delta$ & A \\
36  & 2002-07-15 20:30 & Halo/360 & 1151 & -25.6  & 19:59        & X3.0 & 10030 & $\beta\gamma\delta$ & B \\
37  & 2002-07-16 16:02 & Halo/360 & 1636 & -41.0* &  --          &  --  & 10030 & $\beta\gamma\delta$ & B \\
38  & 2002-08-16 12:30 & Halo/360 & 1585 & -67.1  & 11:32        & M5.2 & 10069 & $\beta\gamma\delta$ & A \\
39  & 2003-10-28 11:30 & Halo/360 & 2459 & -105.2*& 09:51        & X17.2& 10486 & $\beta\gamma\delta$ & A \\
40  & 2003-10-29 20:54 & Halo/360 & 2029 & -146.5*& 20:37        & X10  & 10486 & $\beta\gamma\delta$ & A \\
41  & 2004-07-25 14:54 & Halo/360 & 1333 & 7.0    & 14:19        & M1.1 & 10652 & $\beta\gamma\delta$ & A \\
42  & 2004-11-06 01:31 & Halo/360 & 818  & -81.5  & 00:44        & M5.9 & 10696 & $\beta\gamma\delta$ & A \\
43  & 2004-11-07 16:54 & Halo/360 & 1759 & -19.7  & 15:42        & X2.0 & 10696 & $\beta\gamma\delta$ & A$^{\dag}$ \\
44  & 2005-01-15 06:30 & Halo/360 & 2049 & -30.7* & 05:54        & M8.6 & 10720 & $\beta\delta$ & B \\
45  & 2005-01-15 23:06 & Halo/360 & 2861 & -127.4*& 22:25        & X2.6 & 10720 & $\beta\delta$ & B \\
46  & 2005-01-17 09:54 & Halo/360 & 2547 & -159.1*& 06:59        & X3.8 & 10720 & $\beta\delta$ & B \\
47  & 2006-07-04 21:30 & 199/102  & 308  &  1.6   & 19:06        & C1.4 & 10898 & $\beta$ & A\\
\hline
\enddata
\tablecomments{}
\tablenotetext{a}{~First appearance time in the LASCO/C2 FOV.}
\tablenotetext{b}{~Position Angle/Angular Width given by the LASCO CME catalog.}
\tablenotetext{c}{~Linear speed given by the LASCO CME catalog.}
\tablenotetext{d}{~Acceleration given by the LASCO CME catalog. The superscript `*' indicates that acceleration is uncertain due to either poor height measurement or a small number of height-time measurements.}
\tablenotetext{e}{~GOES soft X-ray flare start time.}
\tablenotetext{f}{~The CMEs under investigation are classified into two groups by two patterns of helicity injection in their source regions: a pattern of monotonically increasing helicity in Group A and a pattern of sign reversal of helicity in Group B. The superscript `$\dag$' indicates the CME events in Group A which occur during a phase of relatively constant helicity after a significant amount of helicity injection. For the detailed description of Groups A and B, refer to \S~\ref{sec3:sub1}.}
\end{deluxetable}

\clearpage

\begin{deluxetable}{cccccccccc}
\tablewidth{0pt}
\tabletypesize{\scriptsize}
\tablecolumns{10}
\tablecaption{Comparison of Different Characteristics between Group A and Group B \label{sec3:sub1:tab1}}
\tablehead{\multirow{2}{*}{Group} & \colhead{CME} & \multirow{2}{*}{$v_{\mathrm{avg}}$\tablenotemark{a}} & \multirow{2}{*}{$a_{\mathrm{avg}}$\tablenotemark{b}} & \colhead{CME} & \multicolumn{2}{c}{CME-related Flare} & \multicolumn{2}{c}{Flux Pattern\tablenotemark{e}} & \multirow{2}{*}{Helicity Pattern} \\
 & Number & & & Type & $F_{\mathrm{avg}}$\tablenotemark{c} & $I_{\mathrm{avg}}$\tablenotemark{d} & $\dot{\Phi}_{avg}>0$ & $\dot{\Phi}_{avg}<0$ & }
  \startdata
  A & 30 & 870 & -24.4 & Single & 1.5 & 0.7 & 63$\,\%$ & 37$\,\%$ & Monotonic increase \\[0.5ex]
  B & 17 & 1330 & -6.3 & Successive & 2.3 & 1.1 & 76$\,\%$ & 24$\,\%$ & Sign reversal \\
  \enddata
\tablenotetext{a}{~Average velocity (km$\,$s$^{-1}$).}
\tablenotetext{b}{~Average acceleration (m$\,$s$^{-2}$).}
\tablenotetext{c}{~Average total time-integrated flux measured in the 1 to 8 {\AA} by the $GOES$ satellite (10$^{-1}\,$J$\,$m$^{-2}$).}
\tablenotetext{d}{~Average flare impulsiveness which is determined from the ratio of peak flux to flare rise time (10$^{-5}\,$W$\,$m$^{-2}\,$min$^{-1}$ ).}
\tablenotetext{e}{~$\dot{\Phi}_{avg}$ is average change rate of total unsigned magnetic flux in a CME source region.}
\end{deluxetable}

\clearpage

\begin{deluxetable}{lcccccc}
\tablecolumns{7}
\tabletypesize{\scriptsize}
\tablewidth{0pt}
\tablecaption{Helicity Injection in 23 Active Regions Producing 30 CMEs in Group A \label{sec3:sub2:tab1}}
\tablehead{\multirow{3}{*}{No.\tablenotemark{a}} & \multicolumn{2}{c}{CME} & \multicolumn{4}{c}{Source Region} \\
& \colhead{$v$} & \colhead{$a$} & \colhead{$\dot{H}_{avg}$\tablenotemark{b}} & \colhead{$\Phi(t_{1})$\tablenotemark{c}} & \colhead{$\dot{\Phi}_{avg}$\tablenotemark{d}} & \colhead{$\Delta\tau$\tablenotemark{e}} \\
& (km$\,$s$^{-1}$) & (m$\,$s${}^{-2}$) & (10$^{40}\,$Mx$^{2}\,$hr$^{-1}$) & (10$^{20}\,$Mx) & (10$^{20}\,$Mx$\,$day$^{-1}$) & (day)}
\startdata
1  			& 523  & -2.9    & 0.4  & 180  & -20 & 2.1 \\
2$^{\dag}$  & 352  & -1.5    & 5.1  & 530  & 61  & 3.3 \\
3 			& 585  &  8.0    & 11.5 & 460  & 14  & 3.0 \\
4  			& 542  & -1.4    & 12.9 & 460  & 15  & 3.2 \\
5  			& 938  & -28.8   & 15.9 & 480  & 19  & 3.6 \\
6  			& 253  & -1.2*   & 0.7  & 120  & -18 & 3.1 \\
7  			& 898  & -23     & 8.4  & 300  & 83  & 1.7 \\
8  			& 444  & -8.7*   & 0.2  & 100  & -13 & 2.5 \\
9  			& 409  &  6.0    & 8.8  & 390  & 30  & 2.1 \\
10 		    & 739  & -7.1    & 23.5 & 560  & -19 & 1.7 \\
11			& 429  & -2.8    & 3.6  & 210  & 7   & 2.8 \\
12			& 641  & -15.5   & 9.7  & 360  & 97  & 0.5 \\
15			& 1674 & -96.1*  & 59.4 & 740  & -45 & 2.5 \\
16 			& 528  & -5.8    & 14.8 & 440  & 3   & 2.7 \\
17 			& 702  &  2.8    & 25.9 & 400  & 23  & 2.7 \\
18 			& 633  & -64.0*  & 23.8   & 450  & 84  & 2.1 \\
19 			& 481  & -10.4*  & 25.8   & 450  & 78  & 1.7 \\
20 			& 1215 & -12.3   & 28.5 & 460  & 66  & 2.8 \\
21 			& 525  & -4.9    & 31   & 530  & -9  & 1.0 \\
22$^{\dag}$ & 798  & -9.8    & 2.1  & 320  & 7   & 3.5 \\
33$^{\dag}$	& 1009 & -14.5   & 8.4  & 620  & -41 & 1.2 \\
34 			& 558  & -25.6   & 30.2 & 810  & 9   & 2.1 \\
35 			& 901  & -0.7    & 40.6 & 810  & 7   & 2.8 \\
38 			& 1585 & -67.1   & 57.8 & 1010 & 80  & 1.1 \\
39			& 2459 & -105.2* & 128.5  & 1440 & -42 & 1.0 \\
40			& 2029 & -146.5* & 104.8& 1470 & -3  & 2.4 \\
41 			& 1333 &  7.0    & 70.6 & 1110 & -10 & 4.6 \\
42 			& 818  & -81.5   & 90.7 & 450  & 77  & 1.1 \\
43$^{\dag}$	& 1759 & -19.7   & 68.5 & 540  & 61  & 2.7 \\
47 			& 308  &  1.6    & 9.9 & 470  & -1  & 2.5 \\
\hline
\enddata
\tablenotetext{a}{~The ID numbers here are corresponding to those in Table~\ref{sec2:tab1}.}
\tablenotetext{b}{~Average of helicity injection rate in the active region under investigation from the start time of helicity measurement $t_{0}$ to the CME occurrence time $t_{1}$.}
\tablenotetext{c}{~Total unsigned magnetic flux in the active region under investigation at the CME occurrence time $t_{1}$.}
\tablenotetext{d}{~The average change rate of the total unsigned magnetic flux during the time period $\Delta\tau$}
\tablenotetext{e}{~The time period between the start time of helicity measurement and the CME occurrence time.}
\end{deluxetable}

\clearpage

\begin{deluxetable}{lcccccc}
\tablecolumns{7}
\tabletypesize{\scriptsize}
\tablewidth{0pt}
\tablecaption{Helicity Injection in 5 Active Regions Producing 17 CMEs in Group B \label{sec3:sub2:tab2}}
\tablehead{\multirow{3}{*}{No.\tablenotemark{a}} & \multicolumn{2}{c}{CME} & \multicolumn{4}{c}{Source Region} \\
& \colhead{$v$} & \colhead{$a$} & \colhead{$\dot{H}_{avg}$} & \colhead{$\Phi(t_{1})$} & \colhead{$\dot{\Phi}_{avg}$} & \colhead{$\Delta\tau$} \\
& (km$\,$s$^{-1}$) & (m$\,$s${}^{-2}$) & (10$^{40}\,$Mx$^{2}\,$hr$^{-1}$) & (10$^{20}\,$Mx) & (10$^{20}\,$Mx$\,$day$^{-1}$) & (day)}
\startdata
13          & 1119 &  1.5    & 2.4  & 830  & -51 & 1.4 \\
14          & 842  &  58.8   & 0.9  & 780  & -51 & 2.4 \\
23 			& 690  &  1.8    & 7.9 & 600  & 22  & 1.7 \\
24 			& 1289 &  2.1    & 5.8 & 610  & 25  & 1.9 \\
25 			& 1245 & -3.3    & 2.2 & 640  & 35  & 2.3 \\
26 			& 1005 & -0.8    & 0.5 & 660  & 38  & 2.6 \\
27 			& 675  & -4.7    & 4.0 & 680  & 40  & 3.1 \\
28 			& 671  & -10.8   & 4.7 & 710  & 41  & 3.5 \\
29			& 495  & -22.9   & 2.5   & 710  & 40  & 3.8 \\
30 			& 980  &  5.8    & 0.5 & 750  & 42  & 4.4 \\
31 			& 1192 &  1.3    & 8.3 & 660  & -8  & 2.5 \\
32          & 2411 &  211.6  & 6.4 & 660  & -9  & 3.0 \\
36 			& 1151 & -25.6   & 7.2 & 900  & 69  & 2.3 \\
37 			& 1636 & -41.0*  & 3.5 & 980  & 75  & 3.1 \\
44			& 2049 & -30.7*  & 5.0 & 810  & 103 & 1.7 \\
45 			& 2861 & -127.4  & 11.1 & 850  & 90  & 2.4 \\
46 			& 2547 & -159.1  & 12.0 & 940  & 79  & 3.8 \\
\hline
\enddata
\tablenotetext{a}{~The ID numbers here are corresponding to those in Table~\ref{sec2:tab1}.}
\end{deluxetable}


\end{document}